\documentclass{PoS}

\def\simleq{\; \raise0.3ex\hbox{$<$\kern-0.75em \raise-1.1ex\hbox{$\sim$}}\; }
\def\simgeq{\; \raise0.3ex\hbox{$>$\kern-0.75em \raise-1.1ex\hbox{$\sim$}}\; }
\newcommand{\eV}{{\rm eV}}

\newcommand{\GeV}{{\rm GeV}}
\newcommand{\TeV}{{\rm TeV}}

\newcommand{\erg}{{\rm erg}}

\newcommand{\kpc}{{\rm kpc}}
\newcommand{\pc}{{\rm pc}}

\newcommand{\cm}{{\rm cm}}
\newcommand{\km}{{\rm km}}

\newcommand{\s}{{\rm s}}
\newcommand{\sr}{{\rm sr}}

\title{Hard Cosmic Ray Sea in the Galactic Center:\\ a consistent interpretation of H.E.S.S. and Fermi-LAT $\gamma$-ray data}

\ShortTitle{Hard Cosmic Ray Sea in the Galactic Center}

\author{D.~Gaggero\\
        GRAPPA, University of Amsterdam, Science Park 904, 1098 XH Amsterdam, Netherlands\\
        E-mail: \email{D.Gaggero@uva.nl}}

\author{\speaker{D.~Grasso}\\
        INFN and Dipartimento di Fisica, Universit\`a di Pisa, Largo B. Pontecorvo 3, I-56127 Pisa, Italy\\
        E-mail: \email{dario.grasso@pi.infn.it}}
        
\author{A.~Marinelli\\
        INFN and Dipartimento di Fisica, Universit\`a di Pisa, Largo B. Pontecorvo 3, I-56127 Pisa, Italy\\
        E-mail: \email{antonio.marinelli@pi.infn.it}}

\author{M.~Taoso\\
        Instituto de F\'isica Te\'orica (IFT), UAM/CSIC, Cantoblanco, Madrid, Spain\\
        E-mail: \email{m.taoso@csic.es}}

\author{A.~Urbano\\
        CERN, Theoretical Physics Department, Geneva, Switzerland\\
        E-mail: \email{alfredo.leonardo.urbano@cern.ch}}

\author{S.~Ventura\\
        INFN and Dipartimento di Fisica, Universit\`a di Pisa, Largo B. Pontecorvo 3, I-56127 Pisa, Italy\\
        E-mail: \email{sofia.ventura@pi.infn.it}}

\abstract{We present a novel interpretation of the gamma-ray diffuse emission measured by H.E.S.S. in the Galactic Center (GC) region and the Galactic ridge. 
Our starting base is an updated analysis of {\tt PASS8} Fermi-LAT data, which allows to extend down to few GeV the spectra measured by H.E.S.S. and to infer the primary CR radial distribution above 100 GeV.
We compare those results with a CR transport model assuming a harder scaling of the diffusion coefficient with rigidity in the inner Galaxy. Such a behavior reproduces the radial dependence of the CR spectral index recently inferred from Fermi-LAT measurements in the inner GP. 
We find that, in this scenario,  the bulk of the Galactic ridge emission can be naturally explained by the interaction of the diffuse, steady-state Galactic CR sea interacting with the gas present in the Central molecular zone.  The evidence of a GC PeVatron is significantly weaker than that inferred adopting a conventional (softer) CR sea. }

\FullConference{35th International Cosmic Ray Conference --- ICRC2017\\
		10--20 July, 2017\\
		Bexco, Busan, Korea}

\begin{document}

\section{Introduction}

In the latest few years the Galactic Center has been the object of an intense high-energy observational campaign.  

The High Energy Stereoscopic System (H.E.S.S.) collaboration recently reported the discovery of a  $\gamma$-ray diffuse emission from a small region surrounding SgrA*  \cite{Abramowski:2016mir}. The emission spectrum is compatible with a single power-law with index $\Gamma_{\rm GC} = 2.32 \pm 0.05_{\rm stat} \pm 0.11_{\rm sys}$ and extends up to $\sim 50 ~\TeV$ with no statistically significant evidence of a cutoff.

A $\gamma$-ray diffuse emission was also measured by  H.E.S.S. from the larger Galactic Ridge (GR) region \cite{Aharonian:2006au}, roughly corresponding to the central molecular zone (CMZ) -- a massive structure rich in molecular gas that extends up to $\sim 250~\pc$ away from the GC along the Galactic plane (GP).  That measurement was very recently updated using 250 hours of data and improved analysis techniques \cite{Abdalla:2017xja}.  
The angular distribution of the events approximately traces that of the gas, apart from the innermost region where a more peaked emission seems to be present.  The spectrum in the whole ridge extends up to $\sim 45 ~\TeV$ and is compatible with a single power law with index 
$\Gamma_{\rm ridge} =  2.28 \pm 0.03_{\rm stat} \pm 0.2_{\rm sys}$, which is in agreement with the slope measured in the inner region surrounding SgrA*.

Assuming that emission to be hadronic, as expected due to the strong losses suffered by electrons in that region, the inferred spectrum of the primary protons is significantly harder than the local CR spectrum measured at the Earth position ($\Gamma_{\rm CR}(r_\odot) \simeq  2.7$ for $E_{\rm CR} > 300~\GeV/{\rm nucleon}$. 
This has been often interpreted as an evidence of a freshly accelerated cosmic-ray (CR) population in that region possibly originated by the supermassive SgrA* black hole or by an intense starburst activity \cite{jouvin2017}. 

Here we explore a different scenario: In our interpretation the largest part of the gamma-ray diffuse emission from the Galactic ridge is originated by the diffuse, steady-state Galactic CR {\it sea} interacting with the massive molecular clouds in the CMZ under the assumption that in that region its spectrum is harder than the local one.
 
This approach is motivated by recent analyses of Fermi-LAT data \cite{Gaggero:2014xla,Acero:2016qlg,Yang:2016jda} showing that the $\gamma$-ray diffuse emission of the Galaxy, and hence the CR primary spectrum, becomes progressively harder approaching the GC along the GP.  
We will use here {\tt PASS8} Fermi-LAT  data to extend down to few GeV the measurement of the $\gamma$-ray diffuse emission spectrum in the CMZ and SgrA* surrounding, showing that this behavior continues down to the inner $\sim 100~\pc$.   
Following \cite{Gaggero:2014xla} we interpret this behavior in terms of a radial dependence of the scaling of the CR diffusion coefficient with rigidity.
We will use the same scenario to compute the CR sea distribution in the GC region and the $\gamma$-ray diffuse emission produced by their interaction with the dense gas in that region and compare those predictions with Fermi-LAT and H.E.S.S. data.  

\section{H.E.S.S. and Fermi-LAT data}

In the first part of this section we report the $\gamma$-ray diffuse emission spectrum from 5 GeV up to 50 TeV in several part of the CMZ region determined from the combination of Fermi-LAT and H.E.S.S. data. 

We extract Fermi-LAT data using the Fermi\ Science Tools {\tt v10r0p5} \cite{Ackermann:2012kna}. We use 470 weeks of {\tt PASS8} data with the event class {\tt CLEAN} and we apply the recommended quality cuts: {\tt (DATA\textunderscore QUAL==1) \&\& (LAT\textunderscore CONFIG==1)}. 
The exposure is computed using the Fermi-LAT response function {\tt P8REP2\textunderscore CLEAN\textunderscore V6}.
Here the data are binned in 20 energy bins equally spaced in log scale between 300 MeV and 300 GeV.
We subtracted the emission due to the point source obtained from the 4-year Point Source Catalog (3FGL) provided by the Fermi-LAT collaboration~\cite{Acero:2015hja}. 

\begin{figure}[t!]
\centering
\includegraphics[width=0.45\textwidth]{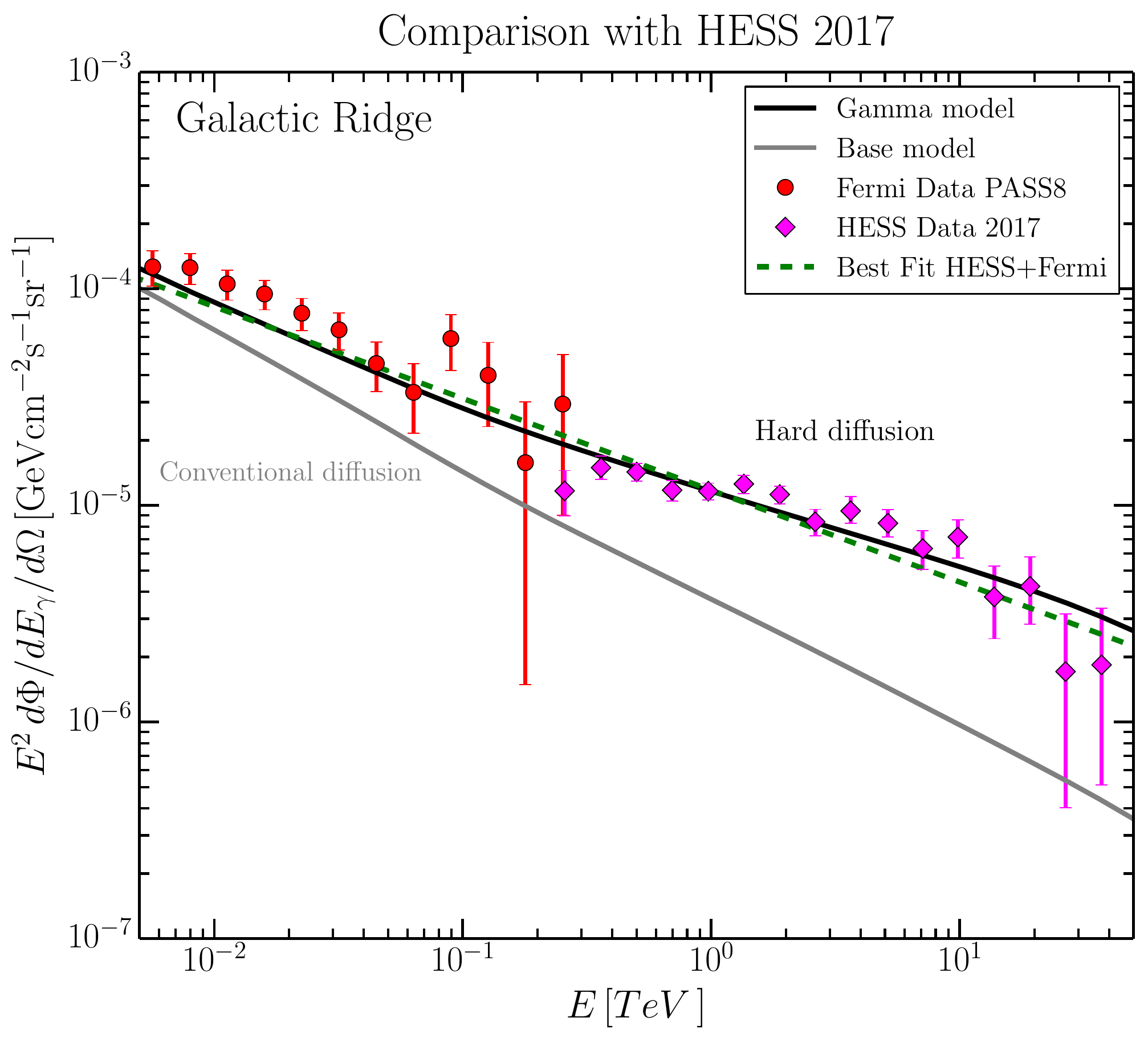}  
\caption{The $\gamma$-ray spectrum in the GR region ($| l | <  1.0^\circ$, $| b | < 0.3^\circ$). Fermi-LAT data, extracted with the  Fermi\ Science Tools in this work, and H.E.S.S. data from \cite{Abdalla:2017xja} are compared with the contribution of the Galactic CR sea as computed with the {\it gamma} and {\it base} models discussed in the text. The single power-law best fit of the combined data is also reported. We have subtracted the contribution of point sources from Fermi-LAT data.
}
\label{fig:GR}
\end{figure}

\begin{figure}[t!]
\centering
\includegraphics[width=0.45\textwidth]{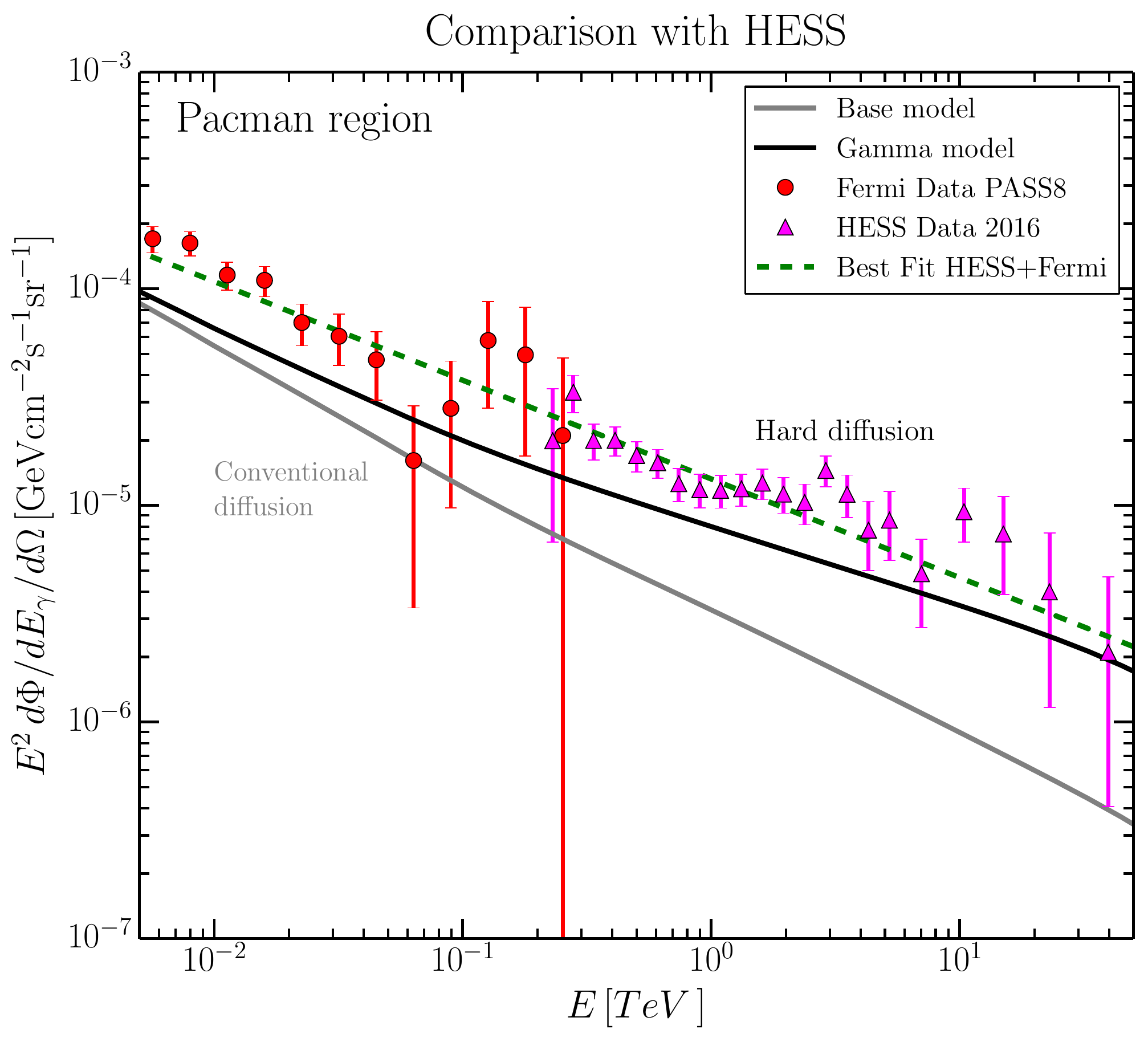}  
\caption{Same as Fig.~\ref{fig:GR} but for the {\it pacman} region defined in the text.
H.E.S.S. data are taken from \cite{Abramowski:2016mir}}
\label{fig:pacman}
\end{figure}

We consider three regions contained in the CMZ complex for which the H.E.S.S. collaboration released the diffuse emission spectra: 
\begin{itemize}
\item the {\em Galactic ridge}  (GR) which in Galactic coordinates is defined by the window  ($| l | <  1^\circ$, $| b | < 0.3^\circ$) and almost include the whole CMZ;
\item the so called {\em pacman, i.e.}  an open annulus centered on SgrA* with $\theta_{\rm inner} = 0.15^\circ$  and $\theta_{\rm outer} = 0.45^\circ$;
\item the Sgr B gas complex ($0.4 < l < 0.9$, $-0.3 < b < 0.2$) which is located at a mean distance of $\sim 100~\pc$ from the GC.
\end{itemize} 
For each of these regions we report the spectra data point in Fig.s  \ref{fig:GR},\ref{fig:pacman},\ref{fig:SgrB} respectively. 

Interestingly, the emission from all these regions can be matched by a single power law with index $\sim 2.4$ and same normalization.
More precisely the best fit spectrum in the ridge region for $5~\GeV < E < 5~\TeV$ is 
\begin{equation}
\Phi_{\rm{GR}} = (1.19 \pm 0.04) \times 10^{-5} \left(\frac{E_{\gamma}}{1~\TeV}\right)^{{-2.42 \pm 0.02}}\left(\GeV~ \cm^{2}~ \s~ \sr \right)^{-1}.
\end{equation}

This finding suggests that a single, almost uniform, emission process may be responsible for most of the emission of the whole CMZ.
The agreement between the Fermi-LAT and H.E.S.S. flux normalization is remarkable taking in mind the very different techniques adopted by those experiments.  

To further investigate the nature of the emission, and similarly to what done from the H.E.S.S. collaboration
we use the angular distribution of Fermi-LAT data to infer the radial distribution of the primary CR energy density $w_{\rm CR}$. 
Using the Fermi tools we extract the diffuse luminosity $L_\gamma(E_\gamma \ge 10~\GeV)$ in an annulus and in six adjacent circular regions with angular diameter of $0.2^\circ$ centered on the plane intersecting SgrA* (see Fig.~\ref{fig:cr_profile}).  
These regions are larger than those considered by H.E.S.S., which is motivated by the smaller angular resolution of Fermi-LAT.
To determine the gas mass distribution we use the same CS column density map adopted by the H.E.S.S. collaboration \cite{Abramowski:2016mir}.
Accounting for the energy dependence of the pion production cross-section we get the following relation between $w_{\rm CR}$ and the $\gamma$-ray luminosity:
\begin{eqnarray}
w_{\rm CR} &&(E_{\rm CR} \ge 0.1~\TeV) = 3.9 \times 10^{-2}~~\eV \cm^{-3} \nonumber \\
 &&\left(\frac{\eta_N}{1.5}\right)^{-1}~\left(\frac{L_\gamma(\ge 10~\GeV)}{10^{34}~\erg/\s}\right) \left(\frac{M_{\rm gas}}{10^6~M_\odot}\right)^{-1}~.
\end{eqnarray}
Here $L_\gamma(\ge E_\gamma)$ is the $\gamma$-ray luminosity above $E_\gamma$ in each region (subtracting the contribution from point sources); $M_{\rm gas}$ is the corresponding total hydrogen mass;
$\eta_N \approx 1.5$ is a factor accounting for the presence of heavier nuclei.   
The resulting CR energy density radial profile $w_{\rm CR}(r)$ in the energy range $0.1 \le E_{CR} \le 0.3~\TeV$ is reported in Fig.~\ref{fig:cr_profile}, as well as the CR distribution derived by the H.E.S.S. collaboration in \cite{Abramowski:2016mir} for $E_{CR} \ge 10~\TeV.$  
Within the large errors and data scatter, both data sets are compatible with a constant CR density for $r \simgeq 100~\pc$. 
At lower radii, however, a density peak centered on the GC seems to be present, although this is more evident in the H.E.S.S. data set.   

\section{The emission due to the CR sea}

In this section we compare our previous results with the diffuse emission due to the interaction of the CR large scale distribution, the CR {\it sea},  with the gas in the CMZ.    
For conventional models, which assume the CR shape in the whole Galaxy to be the same as that measured at the Earth,  the emission was estimated to be considerably smaller and flatter than H.E.S.S. finding. 
Interestingly, below 10 GeV those models are in good agreement with Fermi-LAT data (see e.g. \cite{Yang:2014bcj}), hence 
an excess was found only at very high energies.  This motivated several authors to attribute the excess to a hard CR component freshly accelerated by one or more sources in the GC region. 

Here we consider an alternative scenario in which the Galactic CR spectrum, hence the secondary $\gamma$-ray diffuse emission, shows a progressive hardening at low Galactocentric radii. 
In particular, following \cite{Gaggero:2014xla,Gaggero:2015xza,Gaggero:2017jts}, we reproduce this behavior adopting a CR propagation model featuring a radially-dependent scaling of the diffusion coefficient on the particle rigidity $\delta$. 
The scenario, implemented in the {\tt DRAGON} code \cite{Evoli:2008dv,Evoli:2016xgn}, assumes that $\delta$ has a linear dependence on the Galactocentric radius ($r$): $\delta(r) = A r + B$. The parameters $A$ and $B$ were tuned to consistently reproduce local CR and Fermi-LAT $\gamma$-ray data on the whole sky. 
 
With respect to the model considered in \cite{Gaggero:2014xla} (KRA$_\gamma$), the {\it gamma} reference model considered here   
adopts a spectral hardening in the proton and Helium source spectra at $\sim 300~\GeV/{\rm n}$, in order to reproduce the local propagated spectra measured by PAMELA \cite{Adriani:2011cu}, AMS-02 \cite{Aguilar:2015ooa} and CREAM \cite{Ahn:2010gv}. 
We assume this feature to be present in the whole Galaxy, as it may be expected if it is produced by propagation effects. 
Under these conditions, the KRA$_\gamma$ model was shown \cite{Gaggero:2015xza} to reproduce the emission observed by Milagro in the inner GP at a 15 TeV median energy \cite{Abdo:2008if} consistent with Fermi-LAT data.  

We compute the $\pi^0$, Inverse-Compton and bremsstrahlung components of the $\gamma$-ray diffuse emission, integrating the convolution of the spatially-dependent CR spectrum, gas/radiation density distributions and proper cross-sections along the line-of-sight. The $\pi^0$ component is dominant in the GC region. 
With respect to what reported in \cite{Gaggero:2014xla}, here we replace the hydrogen distribution in the inner $3~\kpc$ with the 3-dimensional analytical model presented in \cite{Ferriere:2007yq}, as required to properly model the hadronic emission in that region.  Outside that region we adopt the gas model used in \cite{Vladimirov:2010aq}.  
The main components are molecular (H$_2$) and atomic (HI) and hydrogen. HI, which is inferred from 21-cm lines, is less than 10\% of the total mass. 
Since H$_2$ is not observed directly, the column density must be inferred from proper tracers, most commonly from the CO emission lines. Here we use a conversion factor $X_{\rm CO}(r \sim 0) \simeq 0.6 \times 10^{20}~\cm^{-2}~{\rm K}^{-1}~\km^{-1} \s,$ the value giving the best agreement with the integrated mass distribution, based on the CS emission map, used in \cite{Abramowski:2016mir}. 

Concerning the CR source distribution by default we use the one reported in \cite{Case:1998qg} based on supernova remnant catalogs. This parametrization vanishes at the GC, a behavior in qualitative agreement with the $\gamma$-ray emissivity profile determined by the Fermi-LAT collaboration \cite{Acero:2016qlg}, which displays a dip in the GC.
We verified that using the source distribution reported in \cite{Yusifov:2004fr}, which does not vanish at the GC, turns into a factor $\sim 2$ larger emission from the GR and {\it pacman} regions. This is still compatible with data and, moreover, may be compensated by a reduction of the $X_{\rm CO}$ factor within the allowed observational uncertainty. 

\begin{figure}[t!]
\centering
\includegraphics[width=0.45\textwidth]{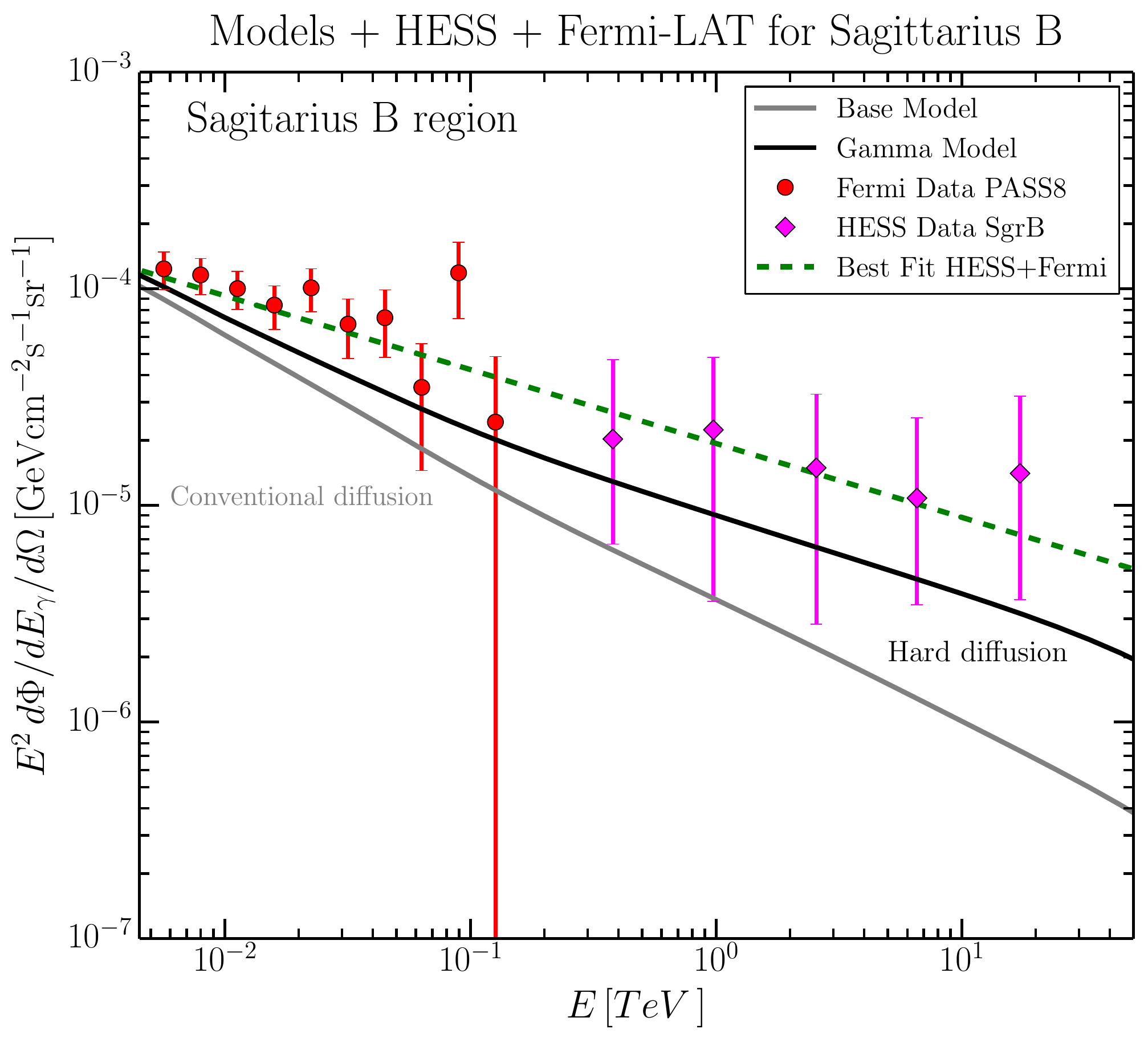}  
\caption{Same as Fig.~\ref{fig:GR} but for the SgrB region defined in the text.
H.E.S.S. data are taken from \cite{Yang:2014bcj}.}
\label{fig:SgrB}
\end{figure}

In the Fig.s  \ref{fig:GR},\ref{fig:pacman},\ref{fig:SgrB} we show, against the experimental data, the $\gamma$-ray diffuse emission spectra due to the CR Galactic sea interaction with the ISM in the Galactic ridge and {\it pacman} and SgrB regions respectively. 
For comparison, besides the prediction of our gamma model we also report the spectra computed for a conventional model ({\it base model}), sharing with the former all the properties but keeping the diffusion coefficient spatially uniform. 
As already well known, from those figures it is evident that the CR sea computed for the base model--as any other conventional model-- cannot consistently account for the H.E.S.S. and Fermi-LAT measurements in the absence of an additional component with a harder spectrum.  

The main novelty of our work is that this conclusion does not hold for models accounting for the radial gradient of the CR spectrum.  
As the reader can see from our figures, the gamma model is in excellent agreement, both in shape and normalization, with those data in the GR and SgrB regions (the latter data set however was taken in a previous H.E.S.S. observational campaign and display large errors). 
Noticeably, the H.E.S.S. data in the ridge region (which were released after the first publication of our results) nicely sit on the model prediction.  

\begin{figure}[th!]
\centering
\includegraphics[width=0.45\textwidth]{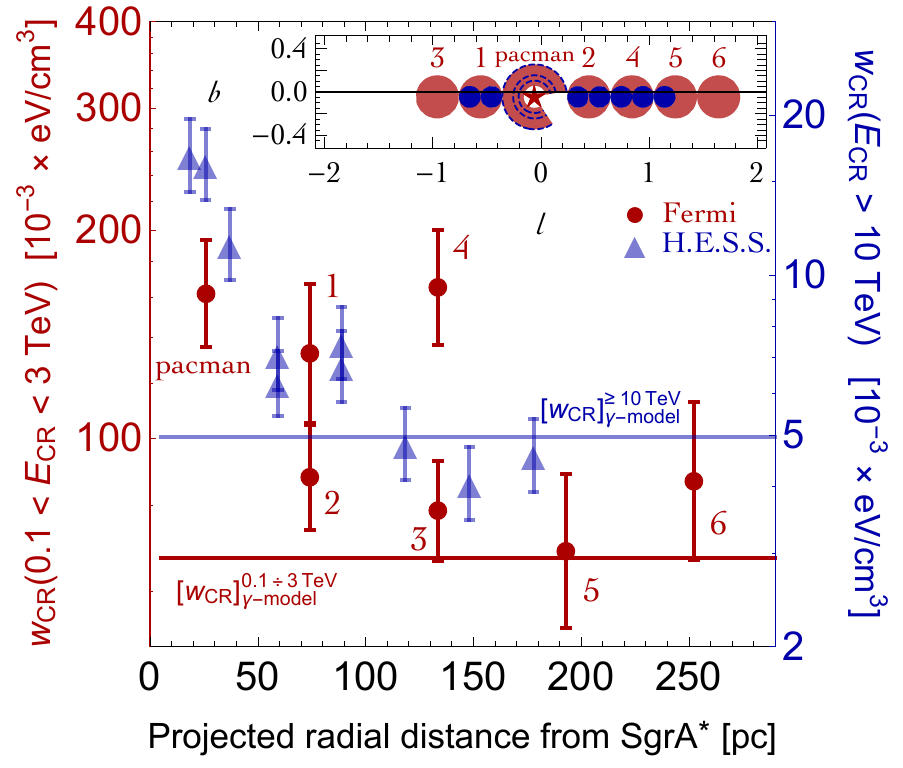}  
\caption{The CR energy density radial profiles for $E_{CR} > 10~\TeV$, as determined by H.E.S.S. \cite{Abramowski:2016mir}, and for $0.1 \le E_{CR} \le 3~\TeV$,  as determined here from Fermi-LAT data, are reported. Those data are compared with the {\it gamma} model predictions (solid lines).
The regions of sky used for deriving the data are represented in the inset.
The model energy density profiles on Galactic scales are reported in the Supplementary material. 
}
\label{fig:cr_profile}
\end{figure}

A small, almost energy independent, deficit with respect to the data is present in the very inner pacman region only.  
This finding is consistent with what inferred from the CR energy density radial profile $w_{\rm CR}$, shown in Fig. \ref{fig:cr_profile} which, compared to the CR sea (almost uniform on scales of few hundred parsecs), displays a peak toward the GC.  
This feature shows up both in the energy range probed by H.E.S.S. and (less significantly) by Fermi-LAT. It could be an artifact due to a gas density underestimation in that region (with respect respect to that inferred by the CO and CS emission maps): It should indeed be noticed that, for gas densities exceeding $10^4~\cm^{-3}$, as expected in that region, the CS and CO emissions should be partially absorbed leading to an underestimation of the gas mass.   
If real, the CR density peak at SgrA* position could be originated by one or more sources close to the GC,  
which are likely to be the same responsible for the J1745-290 emission observed by H.E.S.S. extending up to 10 TeV. 
Although above 100 TeV that emission might be attenuated due to the presence of a dense radiation field around SgrA*, our results show that above that energy the evidence of an excess with respect to the Galactic background is rather small. 

\section{Conclusions}

We have shown that the diffuse $\gamma$-ray emission from the CMZ measured by H.E.S.S. and Fermi-LAT from few GeV up to 50 TeV can be originated by the interaction of the diffuse Galactic CR population with the dense gas present in that region: This implies that a PeVatron at the GC is not required to explain H.E.S.S. data. 

Differently from conventional CR propagation scenarios, we adopted a model based on spatially-dependent diffusion designed to reproduce the radial gradient of the CR spectral index inferred by Fermi-LAT data. 
Therefore, our present results provide a new strong evidence supporting the validity of that setup in a region of the Galaxy were the discrepancies between models featuring standard and radially-dependent diffusion are expected to be maximal.
 
As far as the physical interpretation of this scenario is concerned, we mention that a similar behavior can naturally arise within the framework of anisotropic CR diffusion, given a realistic configuration of the Galactic magnetic field that accounts for a poloidal component in the inner Galaxy \cite{Cerri:2017joy}.  The observed radial trend of the proton spectral index was reproduced under those conditions with the {\tt DRAGON 2} code described in \cite{Evoli:2016xgn,Ligorini:ICRC17}.
 
In the future, the South site of CTA \cite{Acharya:2013sxa} may provide a further confirmation of the scenario discussed in this Letter from the detailed observation of a larger region centered on the GC.

The implications for neutrino astronomy of the results presented in this work are discussed in another talk at this conference \cite{Marinelli:ICRC2017}.

\end{document}